\documentclass[aip,rsi,reprint,graphicx]{revtex4-1} 

\usepackage{graphicx}

\draft 

\begin{document}


\title{Effusive Atomic Oven Nozzle Design Using an Aligned Microcapillary Array} 



\affiliation{Department of Physics and California Institute for Quantum Emulation, University of California Santa Barbara, Santa Barbara, California 93106, USA}

\author{Ruwan Senaratne}
\email[]{Electronic mail: rsenarat@physics.ucsb.edu}

\author{Shankari V. Rajagopal}
\author{Zachary A. Geiger}
\author{Kurt M. Fujiwara}
\author{Vyacheslav Lebedev}
\altaffiliation{Present address: Los Alamos National Laboratory, Los Alamos, New Mexico 87545, USA}
\author{David M. Weld}


\date{\today}

\begin{abstract}
We present a simple and inexpensive design for a multichannel effusive oven nozzle which provides improved atomic beam collimation and thus extended oven lifetimes. Using this design we demonstrate an atomic lithium source suitable for trapped-atom experiments. At a nozzle temperature of $525^{\circ}$C the collimated atomic beam flux directly after the nozzle is $1.2 \times 10^{14}$ atoms per second with a peak beam intensity greater than $5.0 \times 10^{16}$ atoms per second per steradian. This suggests an oven lifetime of several decades of continuous operation. 

\end{abstract}


\maketitle 

\section{Introduction}
Effusive sources have been a common feature of molecular and atomic beam experiments for decades. \cite{ramsey56} Such sources are  useful in trapped-atom experiments on low-vapor-pressure species such as lithium, in which the background vapor pressure is inadequate to trap large numbers of atoms. The capture efficiency of an atomic beam --- the percentage of trappable atoms which enter the trapping region --- is determined in part by the solid angle subtended by the beam. For traditional single-orifice nozzles, this solid angle can be as large as $2 \pi$~sr, leading to a small capture efficiency.  Such nozzle designs necessitate frequent oven changes in order to replenish the supply of atoms, leading to a disruptive periodic breaking of vacuum. 

\begin{figure}
\includegraphics[width=\columnwidth]{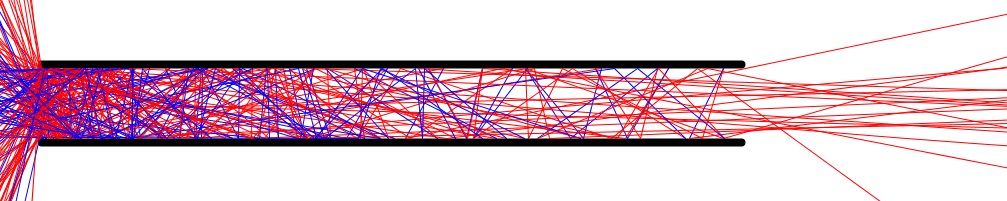}
\caption{(Color online). Simulation of particle trajectories through a 2D tube. In this simple simulation, the particles enter on the left-hand side at uniformly distributed random angles, and assume a random angle upon each collision with a wall. Rightward-moving paths are colored red and leftward-moving paths are colored blue.  Those particles that do not pass all the way through the tube without striking the walls have a higher probability of exiting the tube on the left rather than the right, because they are likely to strike the walls before reaching the midpoint of the tube. Thus the exit beam is collimated, and most particles that would not enter the trapping region are recirculated to the oven reservoir. In this figure the tube aspect ratio has been reduced from 50 to 10 for clarity. \label{simulation}}
\end{figure}

A popular approach to designing high-capture-efficiency atomic and molecular beams is to use nozzles composed of multiple tubes.  If the mean free path is greater than the tube length, the angular spread of the beam is set by the tubes' aspect ratio.\cite{ross95} Most particles that collide with the tube walls exit the tube on the same end that they entered (see Figure \ref{simulation}). In practice, as discussed in Ref. \citenum{ross95}, such nozzles are often operated in an intermediate regime in which the mean free path is less than the tube length but greater than the tube diameter. Operation at very small mean free paths can cause increased angular divergence and depletion of low-velocity atoms. In order to minimize the effects of inter-atomic scattering while maintaining minimal angular spread and high beam intensity, several groups have implemented nozzles consisting of numerous microchannels with diameters ranging from a few microns to a few hundred microns.

Multichannel nozzles were pioneered by Zacharias, who used alternate stacked layers of smooth and corrugated nickel foil to create a matrix of channels.\cite{king56} This technique was also used to form an intense beam of ammonia molecules in the construction of the first maser.\cite{gordon55} A non-exhaustive list of subsequent multichannel designs includes klystron grids, electrolytically etched plastic, photographically etched metal foil and short sections of hypodermic tubing compressed into an aperture.\cite{ross95, giordmaine60, helmer60, johnson66,schioppo12} An alternate approach to the realization of long-lived effusive ovens is to use a recirculating design. Recirculating ovens recover atoms that would not enter the trapping region by cooling them to the liquid state and  returning them to the oven reservoir.\cite{lambropoulos77,swenumson81,drullinger85,hau94,walkiewicz00,slowe05, pailloux07}  However, such designs add substantial complexity with little to no lifetime increase over multichannel designs.

In this paper we present an inexpensive and simple design for a multichannel array consisting of microcapillaries cut from hypodermic tubing. An important challenge in designing such a nozzle is to ensure that all of the microcapillaries are constrained to be parallel to the axis of the experiment. The absence of such constraint has been identified as a limiting factor for multichannel nozzles.\cite{schioppo12} To the best of our knowledge no design that ensures this constraint has been published. 
\section{Design}

In our nozzle we constrain the microcapillaries to be parallel by stacking them in a hexagonally packed lattice. Such packing eliminates the possibility of gaps due to dislocations in the array. This is also desirable, since if such a gap had a width larger than the inner diameter of one of the microcapillaries it could dominate the conductance through the nozzle and thus the solid angle of the resultant atomic beam. 

We enforced the hexagonal packing of the microcapillaries by appropriate design of the bounding walls of the array.  This necessitates that the angle between adjoining boundary walls is either $60^{\circ}$ or $120^{\circ}$. The simplest boundary design forms an equilateral triangle of microcapillaries and requires only one $60^{\circ}$ angle to be machined. Since a large tooling radius at the apex of the triangle would disrupt the ordering of the array, we used wire-cut electrical discharge machining (EDM) for this step. The precision required to ensure no defects in the hexagonally packed lattice over the length scale of a typical nozzle size is well within the capabilities of EDM. 

\begin{figure}[h]
\includegraphics[width=0.8\columnwidth]{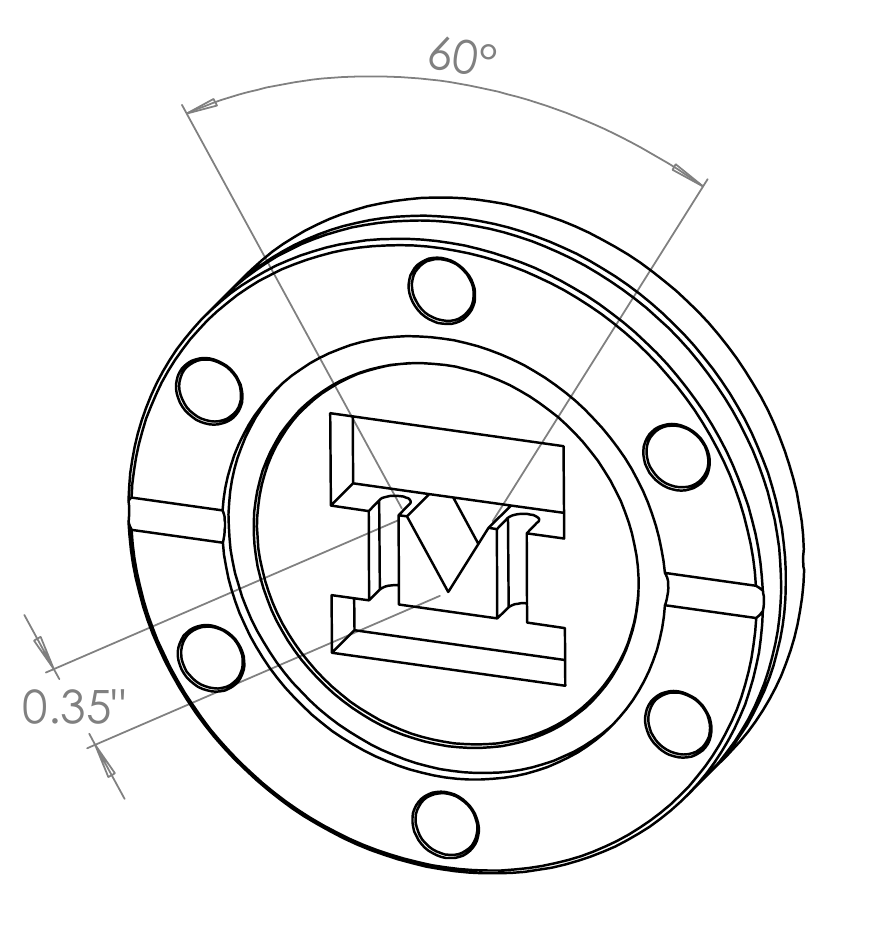}
\caption{Drawing of channel and clamping pocket machined in CF blank. \label{draw}}
\end{figure}

\begin{figure}[h]
\includegraphics[width=1\columnwidth]{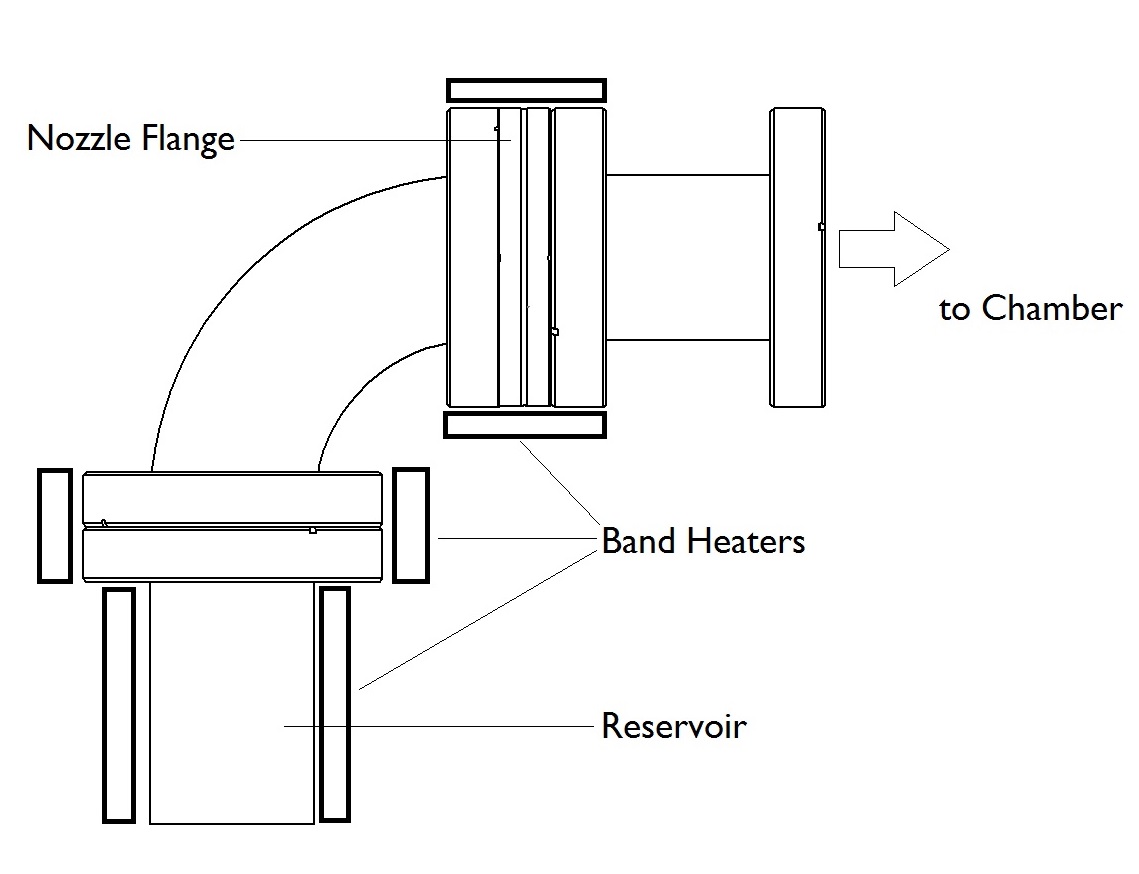}
\caption{Drawing of oven manifold, showing band heater placements.\label{crossection}}
\end{figure}

The nozzle channel is machined out of a $2.75$-inch-diameter double-sided ConFlat (CF) blank, which once assembled, was bolted between two flanges to connect the oven to the rest of the experiment. The nozzle channel is directly machined into the blank to maximize thermal conductance to the exterior, as the  nozzle must be the hottest point in the oven manifold to discourage clogging. The oven manifold is heated using three band heaters, as depicted in Figure \ref{crossection}.  Nickel CF gaskets are used for all seals in the oven manifold.  CF flanges are rated to $450^{\circ}$C, but we have operated such flanges at temperatures over $600^{\circ}$C without any adverse effects on our vacuum. Since the nozzle is machined out of a standard CF  blank, it can be used in an oven manifold for any atomic species that can be safely heated to required pressures within the limits of CF technology. 

\begin{figure}[h]
\includegraphics[width=0.8\columnwidth]{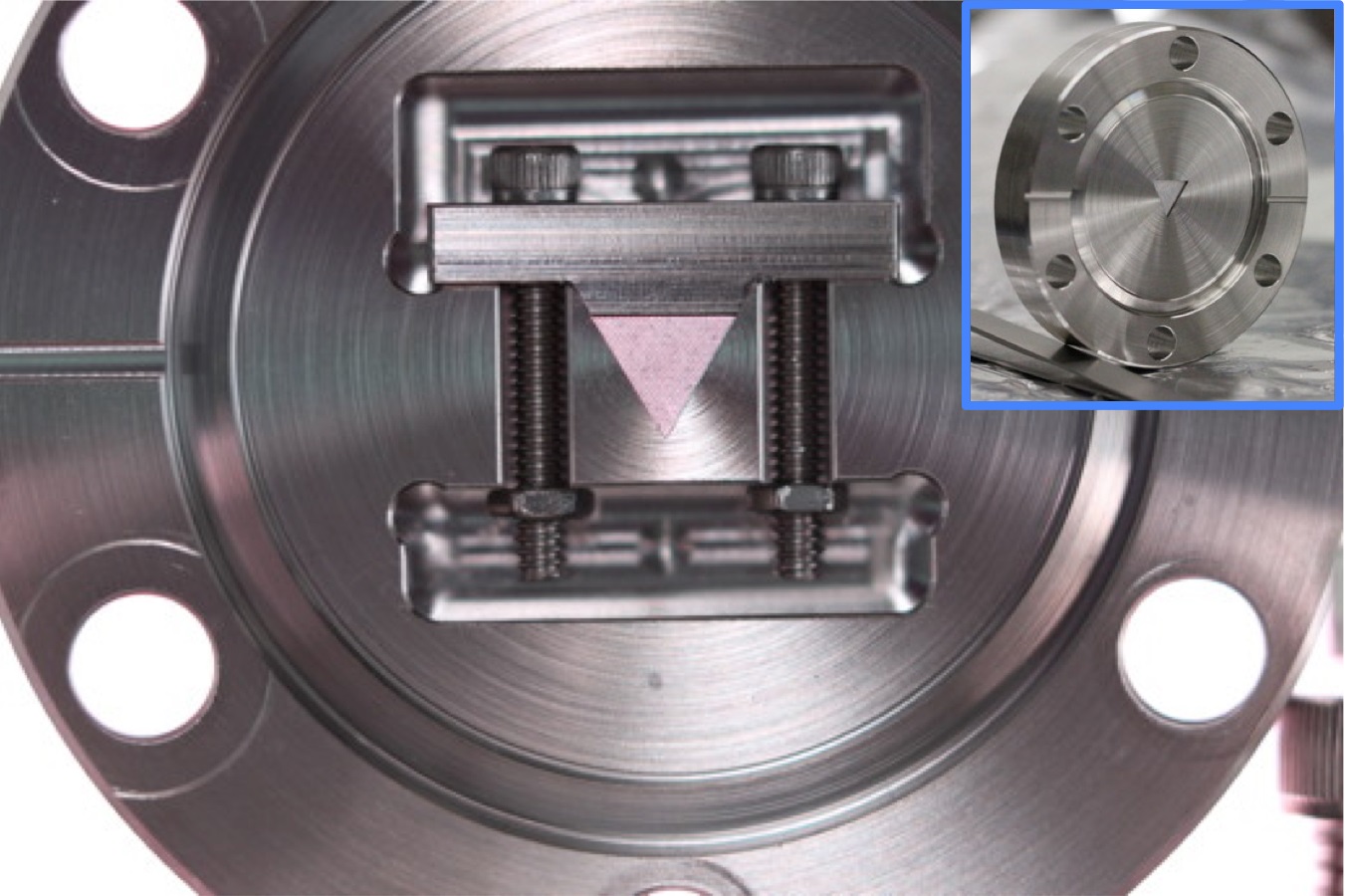}
\caption{(Color online). The assembled nozzle, seen from the side opposite the oven.   Inset shows the oven side of the nozzle.  \label{photo}}
\end{figure}

\begin{figure}[h]
\includegraphics[width=0.8\columnwidth]{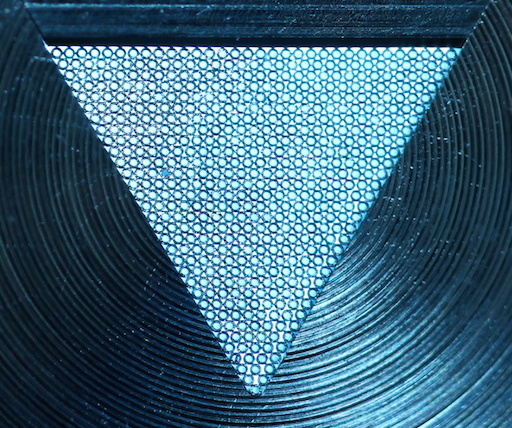}
\caption{(Color online). Close-up of the microchannel array. \label{closeup}}
\end{figure}

The microcapillaries are held in the channel by a single stainless steel clamp that fits into a pocket machined out of the blank. The clamp is precisely sized to firmly hold in 528 microcapillaries in 32 rows, and is tightened with two sets of stainless steel bolts and nuts housed in additional pockets.  These pockets ensure adequate venting for vacuum purposes, so vented screws are not required. The design for the machining of the blank is shown in Figure \ref{draw}, while the completed nozzle is shown in Figure \ref{photo}.  The hexagonal packing of the microcapillaries is evident in the closeup in Figure \ref{closeup}.

The microcapillaries used were 5 mm lengths of 33RW gauge (100 $\mu$m inner diameter) type 304 stainless steel hypodermic tubing purchased from MicroGroup. This length and gauge were selected to ensure a suitable atomic beam solid angle and an atomic flux of the order of $10^{12}$ atoms per second at the trapping region (estimated by calculating the molecular flow conductance).  The solid angle was chosen conservatively to be more than four times the angle subtended by the differential pumping tubes and the Zeeman slower, in order to minimize alignment sensitivity of the 1.7-meter-long machine.

\section{Evaluation}
\begin{figure}[h]
\includegraphics[width=0.9\columnwidth]{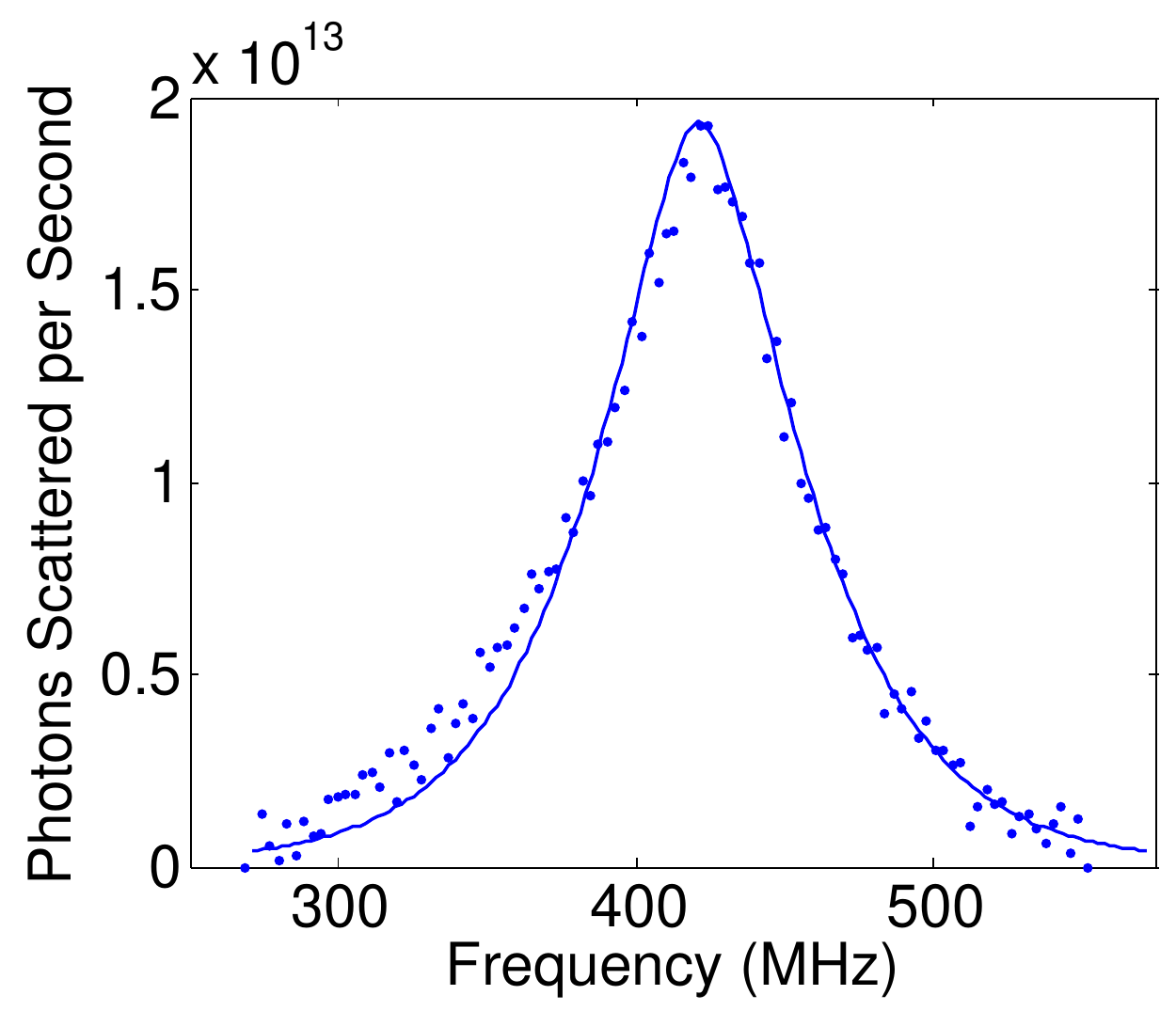}
\caption{(Color online). Measurement of velocity distribution of atomic beam.  Points show photons scattered per second from a resonant transverse laser beam by the atomic beam, at a point three inches downstream of the nozzle. The solid curve shows the expected velocity distibution given the temperature of the oven and the expected divergence angle inferred from the tube aspect ratio. We attribute the asymmetry in the data to an angular deviation of the probe beam from the transverse direction.\label{col}}
\end{figure}

The performance of the nozzle was evaluated in several ways. First, the collimation of the atomic beam and the atomic beam flux were measured approximately 3 inches downstream of the nozzle by recording the absorption of a weak probe laser beam operating on the $^7$Li $2 ^2$S$_{3/2}\longrightarrow 2 ^2$P$_{3/2}$ transition. The probe beam included frequency components resonant with the transitions from both ground-state hyperfine manifolds in order to avoid optical pumping into a dark state. The nozzle was held at a temperature of $525^{\circ}$C for these measurements, with the oven reservoir temperature typically $100^{\circ}$C lower to avoid clogging. No collimating apertures apart from the nozzle itself were present. The results of this measurement are plotted in Figure \ref{col}, which shows the number of photons scattered per second by the center of the atomic beam as the laser frequency was scanned through resonance. The data are consistent with theory for an atomic beam at the nozzle temperature, with a divergence half-angle  in the dimension parallel to the probe beam of $1.2^{\circ}$, set by the tube aspect ratio. Additional absorption measurements were taken at various locations along the axis orthogonal to both the atomic beam and the probe beam in order to measure the angular divergence in this dimension. The measured full-width at half-maximum of the spatial atomic beam distribution in this dimension is $5.8$ mm. Taking into account the known diameter of the nozzle, this value is in accord with the same divergence half-angle of $1.2^{\circ}$.  These measurements demonstrate that the collimation of the atomic beam is controlled by the aspect ratio of the microcapillaries, despite the fact that the calculated mean free path is in the intermediate range between the tube diameter and tube length.   Physically, this can be explained by noting that the decrease in density along the length of the microcapillaries will lead to a substantial increase in mean free path, suggesting that interparticle interaction effects might not become important until the mean free path in the oven approaches the tube diameter.   

These absorption data were also used to calculate a total atomic beam flux of $1.2\times 10^{14}$ atoms per second.  However, our simulations indicate that the angular distribution of atoms from the high-aspect-ratio tubes has  broad non-gaussian shoulders in addition to the sharp central peak.  Including the effects of these shoulders, we  estimate an oven lifetime of over $50$ years of continuous operation at these temperatures given an initial oven reservoir of 25 g of lithium.  As an alternative measure of efficiency, we calculate that the nozzle consumes roughly 200 times less lithium than a hole of the same area to produce a given beam flux in the central peak.  We have observed no performance change in the atomic beam over nine months of operation, which suggests no clogging of the nozzle or unanticipated depletion of the reservoir. 

\begin{figure}[h]
\includegraphics[width=0.90\columnwidth]{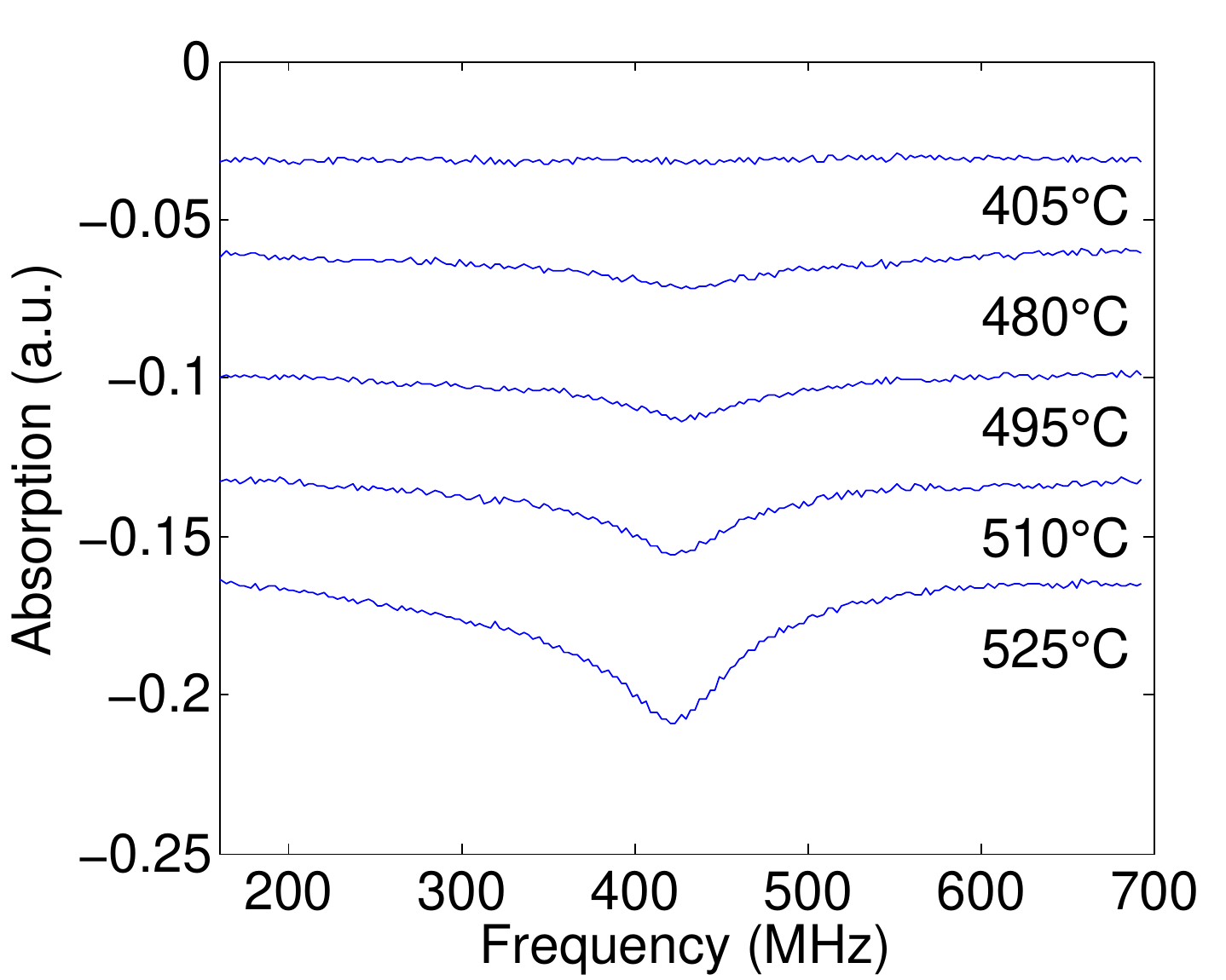}
\caption{(Color online). Measurement of angular distribution and temperature dependence.  Absorption versus frequency of a transverse probe beam by the atomic beam at various nozzle temperatures. Data at different temperatures are offset for clarity.\label{temp}}
\end{figure}

Second, we investigated the performance of the oven nozzle at different temperatures. The absorption versus frequency of a transverse probe laser when aimed through the center of the atomic beam was measured as a function of nozzle temperature, up to $525^{\circ}$C. Angled-beam velocimetry measurements indicate that the temperature of the atomic beam is close to the nozzle temperature.  Transverse-beam data are plotted in Figure \ref{temp}, and show a monotonic increase in atomic beam flux, further supporting the conclusion that stalling is not yet important at these temperatures.

Finally, in conjunction with a Zeeman slower and a standard transverse cooling stage \cite{joffe93}, the atomic beam was used to load a magneto-optical trap (MOT) of up to $1.2 \times 10^9$ atoms, located 1.7 m from the oven, at an initial loading rate of $3 \times 10^8$ atoms per second. 

\section{Conclusion}
We have presented a simple and inexpensive method to implement a multi-tube effusive oven nozzle. The design ensures that all microcapillaries are parallel, eliminates stacking defects, is adequately vented, ensures sufficient thermal conductivity for effective heating, and can be adapted for other species. The nozzle produces a highly collimated lithium atomic beam of $1.2\times 10^{14}$ atoms per second with beam intensities over $5.0 \times 10^{16}$ atoms per second per steradian while maintaining an expected oven lifetime of several decades.


%
%

%

\begin{acknowledgments}

The authors acknowledge support from the Army Research Office and the PECASE program (award W911NF-14-1-0154), the Air Force Office of Scientific Research (award FA9550-12-1-0305), the Office of Naval Research (award N00014-14-1-0805), and the Alfred P. Sloan foundation (grant BR2013-110).  SR acknowledges the National Science Foundation for support via the graduate research fellowship program.  All authors thank Eric Corsini for taking the photographs used in this paper. 

\end{acknowledgments}


%

\end{document}